\documentstyle[twocolumn,prc,aps]{revtex}
\input epsf
\newcommand{\be}{\begin{eqnarray}}
\newcommand{\ee}{\end{eqnarray}}

\begin{document}

\twocolumn [ 
  \hsize\textwidth\columnwidth\hsize\csname @twocolumnfalse\endcsname
  \title {
    { An unusual space-time evolution for  heavy ion collisions 
           at high energies  \\ due to the QCD phase transition
    } 
  }
  \author {
     D. Teaney and E.V.~Shuryak
  }
  \address {
     Department of Physics and Astronomy, State University of New York, 
     Stony Brook, NY 11794-3800 
  }
  \date{\today}
  \maketitle    

\begin{abstract}
The space-time evolution of high energy $non$-$central$ 
heavy ion
 collisions is studied  with relativistic hydrodynamics.  The results are
very sensitive to the Equation of State
(EoS). For an EoS with the QCD phase transition, 
an unusual matter distribution develops.
Before freeze-out, two $shells$ are formed which 
physically separate and leave a maximum in the center. 
We make specific predictions 
for the azimuthal dependence of the flow
and for two-pion interferometry,  contrasting our results
with a resonance gas EoS.
\end{abstract}

\vspace{0.1in}

]

\begin{narrowtext}   
\newpage 
1. One of the principal goals of the heavy ion collision program is to find
and to quantify  the QCD phase transition from hadronic matter 
to a new phase, the quark-gluon
plasma (QGP) \cite{Shu_80}. 
Experiments at the Brookhaven AGS 
(lab energy 11 A*GeV) and at the CERN SPS (lab energy 200 A*GeV) 
are expected to produce a QGP/mixed phase during the initial stages
of the collision, although currently 
there is only indirect evidence for this state
(see the recent reviews \cite{sig-review}).
With the completion of 
the Relativistic Heavy Ion Collider(RHIC) at Brookhaven
and its much higher collision energy (100+100 
GeV*A in the center of mass frame), 
the experiments are expected to produce the QGP 
well above the transition temperature. 
In this work, we study
how the strong QGP pressure can be observed at RHIC.

The  position-momentum correlations
of the produced hadrons, colloquially known as collective $flow$,
directly reflect 
the EoS of the excited matter.
 Multiple  studies
  using cascade event generators 
and hydrodynamics (see e.g. \cite{Pasi,Sorge-RQMD,HS-freeze} ), 
have successfully reproduced the AGS/SPS hadronic
spectra. A radial flow velocity of about (0.5-0.6)c is found in central 
PbPb collisions \cite{sig-review},  but the flow 
develops principally during the 
late hadronic stages of the collisions and has little to do with the
QGP.
An EoS extracted from 
these model studies shows ``softness'' during
the early stages of the collision,
either due to the proximity of the QCD phase transition 
\cite{HS-freeze}, or due to
non-equilibrium phenomena such as the formation and
fragmentation of strings \cite{Sorge-pre}.

2. Additional information about the EoS may be extracted 
from the azimuthal dependence of flow in {\it non-central collisions}, 
which depends non-trivially on the impact parameter and the collision energy.
The ellipticity of the flow  
has been  studied 
theoretically\cite{Ollitrault,Sorge-kink,Sorge-elliptic,Danielewicz}
and experimentally \cite{AGS,Na49}. 
Because elliptic flow develops earlier than radial flow, 
its systematic measurement at the SPS may settle the mixed
phase/pre-equilibrium controversy mentioned above.
 The original purpose of the present study was to further quantify 
elliptic flow within a hydrodynamic framework.
Instead, we found that non-central collisions at RHIC/LHC energies 
have an unusual expansion pattern, which cannot
be described as simply elliptic  and which is
qualitatively different from AGS/SPS energies.

3. Let us begin with a description of the transverse acceleration. 
In model calculations, the  radial acceleration history changes 
 from SPS to RHIC due to the QCD phase transition\cite{KRLG}. 
The ratio
of pressure to energy density, $p/\epsilon$, has a deep minimum at the
end of the mixed phase, known as the
``softest point" of the equation of state 
\cite{HS-soft}. 
 For AGS/SPS collision energies  the matter is produced
 close to the softest point and  
the resulting transverse acceleration is small. 
Therefore, in  non-central collisions the matter retains 
its initial elliptic 
shape and burns slowly inward. 
For RHIC/LHC collision energies, the early pressure
starts an outward expansion\cite{KRLG}.  This outward 
expansion and the inward deflagration can cancel each other,
making a stationary front, called the
``burning log" in \cite{Rischke-log}.
Summarizing,
at the AGS/SPS there is first softness and then a hadronic
push, while at RHIC/LHC  there is first a quark-gluon push,
then softness, and then a hadronic push. 
In spite of this change,
the final radial flow velocities at RHIC and at the SPS
are expected to be similar \cite{HydroUrqmd}.

4. The early push redistributes the matter, however. 
The early velocity
has long time, $\sim 10 fm/c $, to influence the matter 
distribution before  freeze-out.
The stiff QGP in the center, with $T>>T_c$ , pushes against the soft 
matter on the exterior, with $T\approx T_c$,   producing a shell-like 
structure. Since the final distorted distribution rather resembles a nut
and its shell,  we call this picturesque configuration the 
$nutshell$.  
  For high energy {\it non-central} collisions,
the matter expands preferentially in the  impact parameter direction 
(the x axis) and the expanding shells 
leave a  rarefaction behind.
Furthermore, since
the acceleration
started rather early, the two half-shells partially {\it separate},
 and by freeze-out three distinct fireballs are
actually produced. We have called this consequence of 
early pressure the $nutcracker$ scenario.

5.
 Following \cite{Bjorken,Ollitrault},  we assume a rapidity-independent 
 longitudinal expansion.
We then solve the 2+1 dimensional   
relativistic Euler equations in the transverse plane, 
with the coordinates x,y and proper
time $\tau=(t^2-z^2)^{1/2}$, using 
the HLLE Gudunov method \cite{Schneider}. 
As in  previous calculations \cite{HS-soft},  
we have used a simple bag model equation of state 
with $T_c=160 MeV$ and a $1 GeV$ latent heat. We
have modeled hadronic matter with a simple resonance gas
EoS, $p=.2\epsilon$ \cite{resonancegas}.
The pressure was taken \linebreak 
%
%
\begin{figure}[t]

  \vspace{-.2in}
  \epsfxsize=2.60in
  \vspace{-.1in}
  \epsfxsize=2.6in
  \centerline{\epsffile{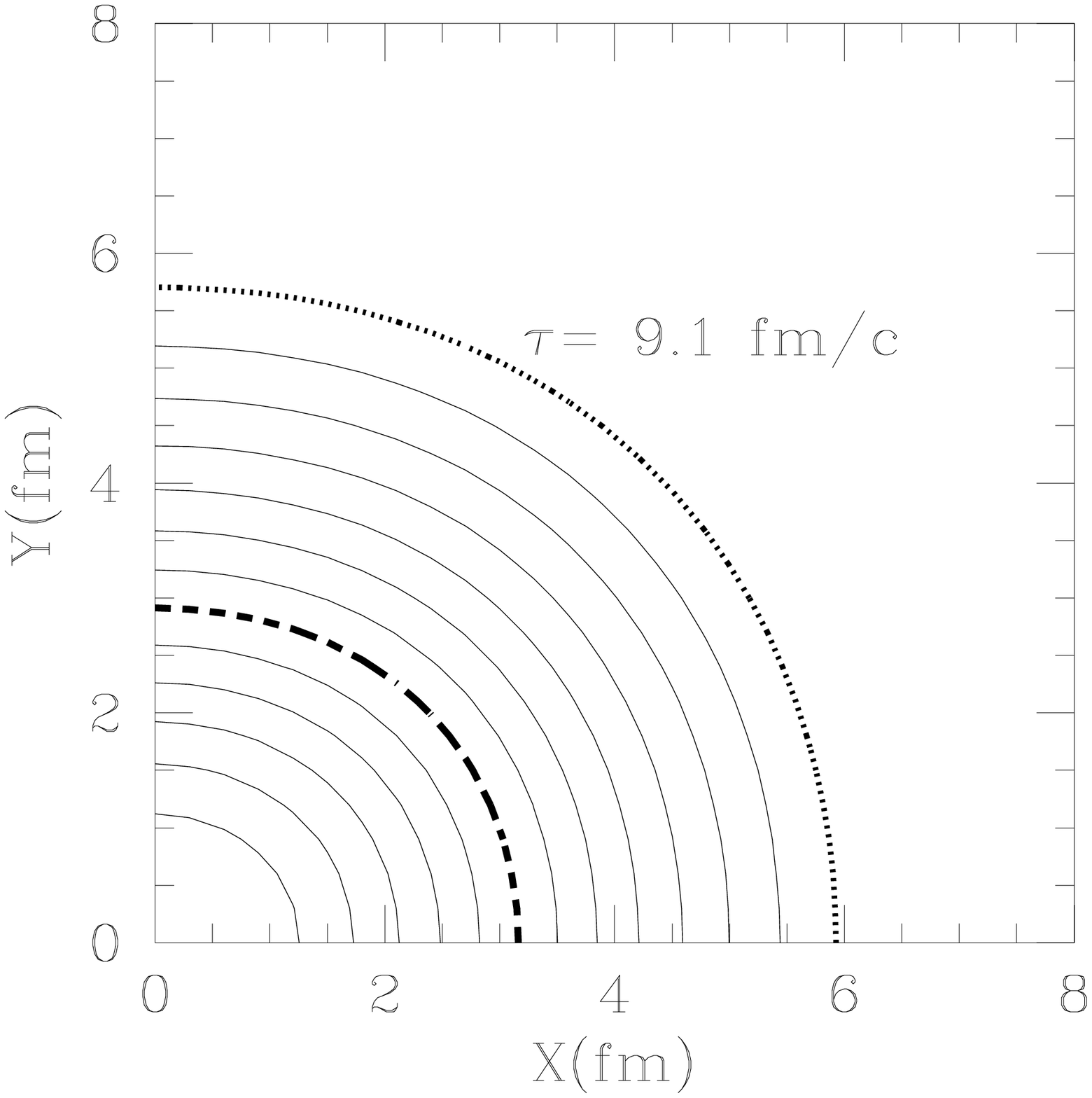}}
  \vspace{-.1in}
  \epsfxsize=2.6in
  \centerline{\epsffile{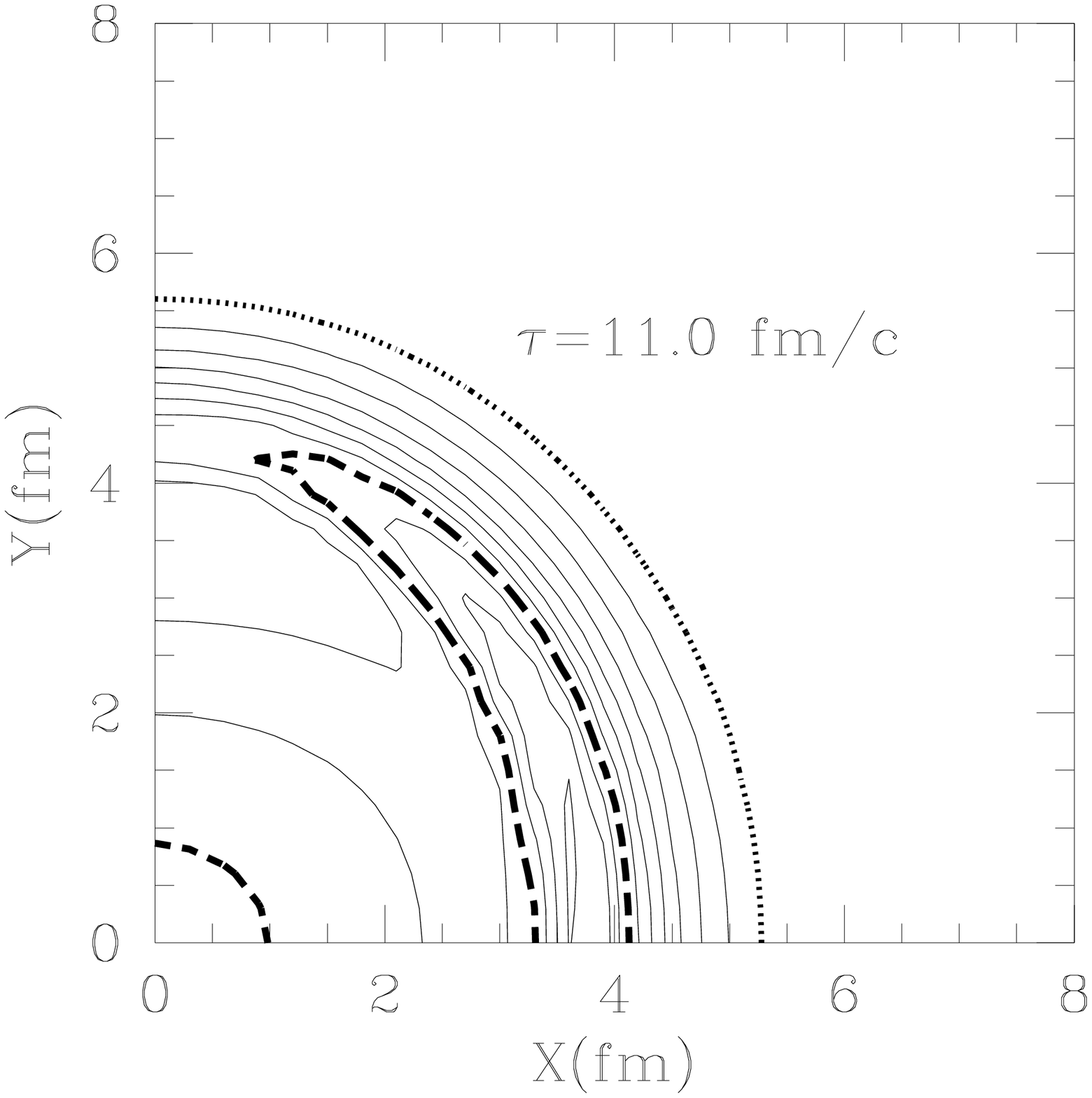}}
  \vspace{.05in}
  \caption[]
  {
   \label{nut-fig}
	Typical matter distributions in the transverse plane
	at mid rapidity, 
	calculated using boost invariant hydrodynamics, 
	for PbPb collisions at b=8 fm and
	 proper time $\tau$ in a representative RHIC collision.
	For the corresponding central collision 
	the pion multiplicity, including all isospin states, was 
	$dN_{\pi}/dy=850$.
      The outer dotted line and inner dashed line  show 
	temperatures of 120 and 140 MeV respectively.
	The solid lines show contours of constant energy density
	with a step size of 10 MeV.  
	(a) shows the distribution at RHIC for a resonance
	gas EoS, p = .2$\epsilon$.
	(b) shows the distribution at  RHIC for a
	bag model EoS. 
   }
\end{figure}
%
%
\noindent to  be independent
of baryon number which is a good approximation at high energies. 
The initial entropy distribution in the transverse plane 
was assumed to be proportional to distribution
of participating nucleons as in \cite{Ollitrault}.
We parameterize the initial energy density
by the total pion multiplicity, $dN_{\pi} / dy$.  
 (How  particle multiplicity maps 
 to the collisions energy depends on the
entropy production mechanism. This mapping will soon be
determined experimentally
  at RHIC.)
For definiteness,  we consider PbPb collisions 
at an impact parameter of $b = 8 fm$  and freeze-out
at a fixed temperature,  $T_f = 140 MeV$.

At SPS energies the flow develops late, 
and the matter retains
its initial almond shape until the late hadronic stage.
However, at RHIC the flow develops early and \linebreak
%
%
\begin{figure}[ht]

  \vspace{-.2in}
  \epsfxsize=2.60in
  \vspace{-.1in}
  \epsfxsize=2.60in
  \centerline{\epsffile{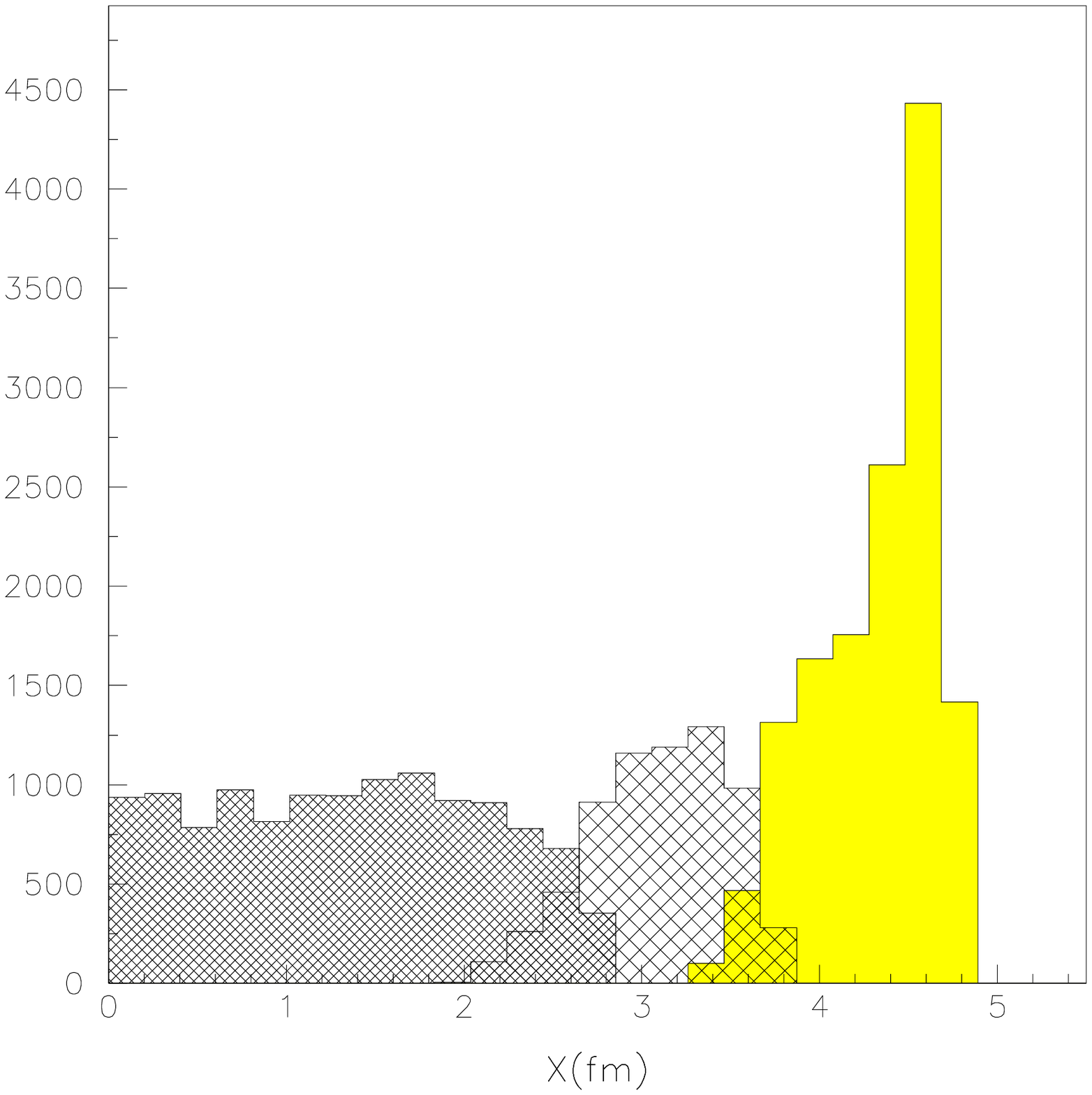}}
  \vspace{-.1in}
  \epsfxsize=2.60in
  \centerline{\epsffile{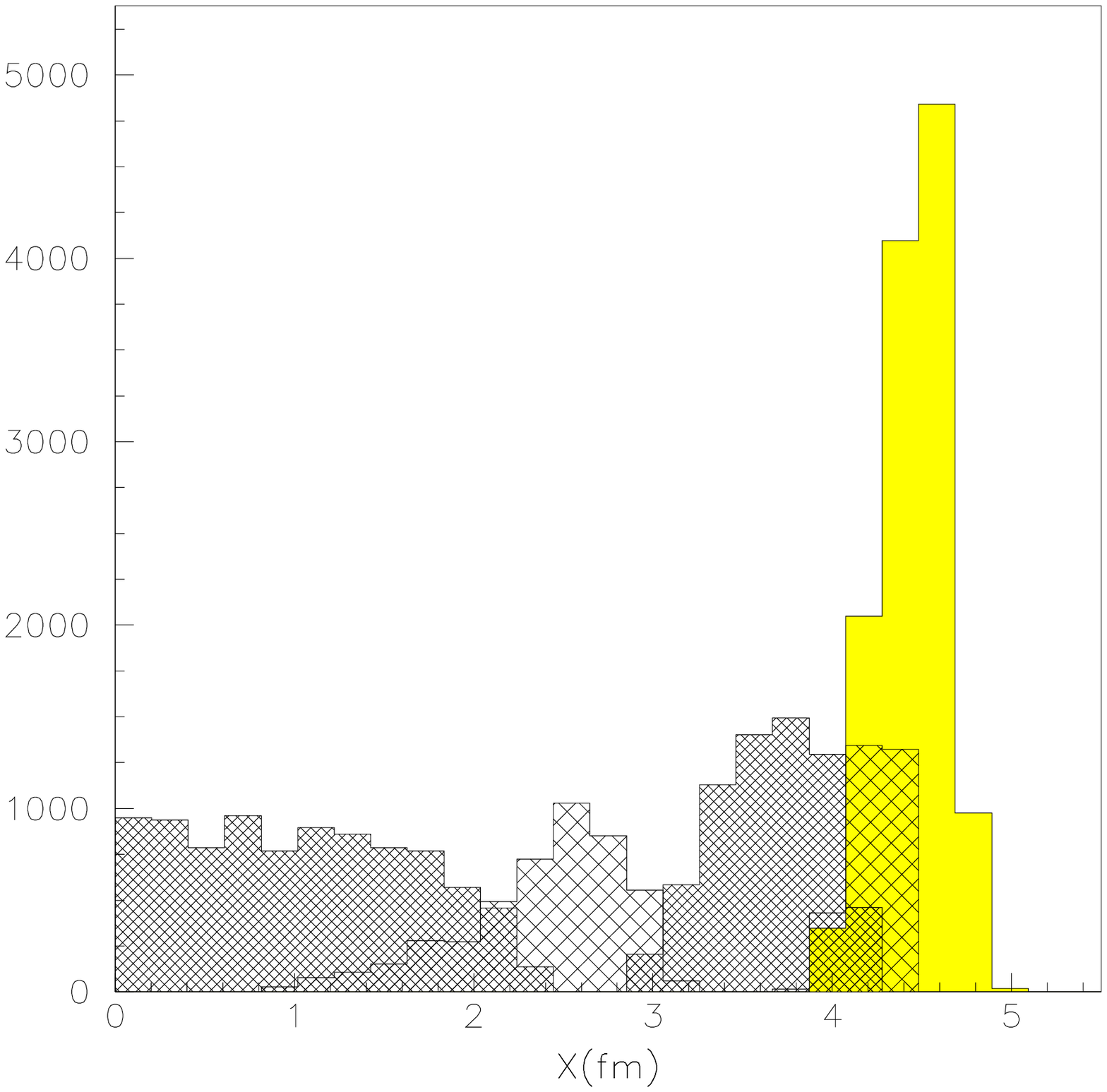}}

  \vspace{.05in}
  \caption[]
  {
   \label{freeze}
The distribution (in arbitrary units) of the x coordinates 
of nucleons emitted from the spatial band, $|y|<2 fm$, and
integrated over intervals of proper time.	
	(a) shows the distribution  for a resonance
	gas EoS. Grey, hatched and dark histograms are
        for $\tau<8$,  $8<\tau<9$ and $9<\tau<(\tau_{max}=10.5)$ fm/c
	  , respectively. 
	(b) shows the distribution for a bag model EoS. 
  	Grey, hatched and dark histograms are
      for $\tau<10$,  $10<\tau<11$ and $11<\tau<(\tau_{max}=12.5)$ 
	fm/c, respectively. 
  }

\end{figure}
%
%
\noindent redistributes 
the matter by late the hadronic stage.
Two sample matter distributions, in the transverse plane and
at fixed proper time,  are shown in Fig. \ref{nut-fig}.
A resonance gas EoS,  $p=.2\epsilon$, produces 
little structure and simple elliptic flow (Fig. \ref{nut-fig}a).
An ideal gas EoS, $p=\epsilon/3$,  also produces simple elliptic flow
and even shorter lifetimes (not shown).
Finally,  our bag model EoS produces
two ``nut-shells'' of matter which expand outward 
(Fig. \ref{nut-fig}b). 
Note that in Fig. \ref{nut-fig}b  the
 matter is pushed into two shells, moving in x direction, while
at the north and south poles, two holes develop. A
maximum, the ``nut'',  remains in the center.
The matter distribution becomes increasingly
``nutty'' with larger impact parameters and higher collision energies.

The evolution is clarified by plotting 
the emission points of nucleons, integrated over periods of proper
time. In Fig.\ref{freeze}, (a) and (b), 
we show the x coordinates of the emitted
nucleons for a resonance gas EoS and for a bag model EoS respectively. 
For both EoS, during the early stages, particles are slowly emitted from   
from a stationary freeze-out surface, making a peak at $x=4-5$ fm 
\cite{Rischke-log}.
For both EoS, however, 75\% of the nucleons freeze-out
during a short proper time interval of 2.5 fm/c. 
For a resonance gas EoS
(Fig. \ref{freeze}a) the freeze-out positions are uniform 
and become increasingly centralized with time. In contrast, for
a bag model EoS (Fig. \ref{freeze}b) the distribution has three distinct
and comparable sources for the final 1.5 fm/c. There 
is a central ``nut'' and two extremal ``shells'' with a hole 
in between. 

6. We turn now to the experimental consequences of this flow
pattern. First we
	examine the $\phi$ distribution of the produced particles
at mid-rapidity (y=0).
These distributions
 are expanded  in harmonics and are
sometimes weighted by the transverse momentum squared.
\be
	{\left. \frac {dN} {d\phi dy}  \right| }_{y=0} 
	= {v_0 \over 2\pi}(1 + \sum_{n\geq1} 2 v_{2n} \cos(2n\,\phi)\,) \\
       \int p_t^2 {\left. \frac {dN} {dp_t d\phi dy} \right| }_{y=0} dp_t 
	= {\alpha_0 \over 2\pi}
	(
	1 + \sum_{n\geq1} 2 \alpha_{2n} \cos(2n\,\phi)
	\,) 
\ee
\noindent 
We have calculated
the single particle distributions for 
various secondaries,
using the standard Cooper-Frye formula \cite{Cooper}.
The elliptic components, $v_2$ and $\alpha_2$, depend 
only weakly on collision energy, as found
in previous studies \cite{Ollitrault}. For nucleons,
for example, we found $v_2 \approx 7\%$ and $\alpha_2 \approx
13\% $ from the highest SPS energies to LHC energies.

Higher harmonics, in contrast, grow from SPS to RHIC. To 
summarize the effects of higher harmonics 
in the distributions,  we have plotted in Fig. \ref{moments-fig} 
the weighted net nucleon 
$\phi$ distribution (the l.h.s. of equation(2)) for SPS and RHIC, normalized to the first two terms 
in the Fourier
expansions shown above. In 
the dashed curve corresponding 
to RHIC,
the  early pressure forces
a $3-4\%$ additional asymmetry 
in final net nucleon distributions 
{\it beyond} the   elliptic component. 
At LHC energies
the additional asymmetry is even more pronounced.
The marked minimum at 45$^o$ is due to the square shape of the mater
distribution, which somewhat reduces the flow on the diagonal. Note also a
prominent positive correction to elliptic flow at 90$^o$. Both of these effects
are  observable, given the expected statistics at RHIC.

The distribution of deuterons and  other heavy fragments should express
the underlying flow more clearly.
Because the emission points
of the nucleons are bunched along the ridges of the nutshell,
larger fragments are 
generally emitted from the shells.
	This inhomogeneity  enhances the
production probability of fragments and peaks 
their final flow in the x direction.
Multiply strange barons such as $\Omega^-$ are also of interest. 
Since they do not re-scatter in the hadron phase,
they reflect the early flow 
\cite{Sorge-strange}. 
Indeed any azimuthal dependence of the flow 
of multiply strange baryons
would be fairly convincing evidence of collective motion in
the quark phase.
%
%
\begin{figure}[h]

  \epsfxsize=2.6in
  \centerline{ \epsffile {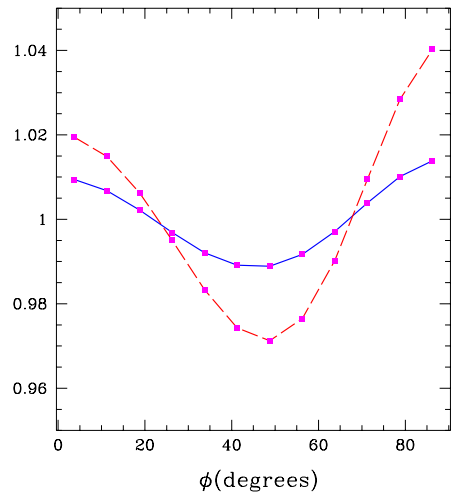 } }
  \vspace{.08in}
  \caption[] 
  { 
   \label{moments-fig}
    The net nucleon $\phi$  distribution 
    weighted by $p_t^2$  (the l.h.s. of equation(2)) 
    and divided by  the first two
    moments of the corresponding Fourier expansion
    (the r.h.s. of equation(2)). This 
    plot summarizes the effect of all higher moments. The
    solid curve is for the SPS and 
    the dashed curve is for RHIC. 
    In the corresponding central SPS and RHIC collisions 
    the pion multiplicity, 
    including all isospin states, was $dN_{\pi}/dy=525$
    and $dN_{\pi}/dy=850$ respectively.
    We summarize the numerical values below.\linebreak
\begin{tabular} {llccr}
    At the SPS&:&$<p_t^2>$=.67 $GeV^2, \,$ & $\alpha_2=12.0 \%, \,$ & $\alpha_4=0.6\%$ \\  
    At RHIC   &:&$<p_t^2>$=.84 $GeV^2, \,$ & $\alpha_2=13.8 \%, \,$ & $\alpha_4=1.3\%$ \\
\end{tabular}
    Note that $\alpha_0$ is proportional to $<p_t^2>$.
  }

\end{figure}
%
%
7. The $spatial$ asymmetry of the matter distribution at freeze-out
is probed by Hanbury Brown-Twiss (HBT)
two particle interferometry. 
	Strong flows strongly modify the source function. Each correlator
with given pair momenta, is
generated by its own ``patch'' , or ``homogeneity region'' \cite{MS}.  
Taking these patches together gives a complete
picture of the source.      
We will discuss this complicated issue
elsewhere, and here show only two selected correlators which
emphasize the qualitatively different predictions 
of different EoS.

The correlators are found by taking
the appropriate Fourier transform of the source function over the 
freeze-out surface \cite{Cooper}. Below we display the
Hanbury Brown-Twiss(HBT) radii $R_{xx}$ and $R_{yy}$ and employ the notation 
$(p_1 + p_2 )_{\mu} = 2K_{\mu}$
and 
$(p_1 - p_2 )_{\mu} = q_{\mu} $. For $R_{xx}$ we select 
 $\vec{q} $  in the x direction , the direction we want to probe. 
 $\vec{K}$ is chosen in the orthogonal y direction , 
 with magnitude .5 GeV. 
For $R_{yy}$ the axes of $\vec{K}$  and $\vec{q}$ are simply reversed.
The correlators may then be fit to the
functional form $C = 1 + exp( - q_i^2 R_{ii}^2) $, where $R_{ii}^2$
has been interpreted as the source size at zero velocity
\cite{Uli-review}. More
specifically,   
\be
	 R_{xx}^2 =  < x^2 > - < x >^2\\ \nonumber
	 R_{yy}^2 =  < y^2 > - < y >^2 \nonumber
\ee
These radii are shown in Fig. \ref{hbt-fig} for
PbPb collisions at b=8 fm, as a function of the pion 
multiplicity  scaled by the number of participants 
to central collisions (not b=8 
\linebreak[4]
%

\begin{figure}[tb]

  \epsfxsize=2.9in
  \centerline{\epsffile{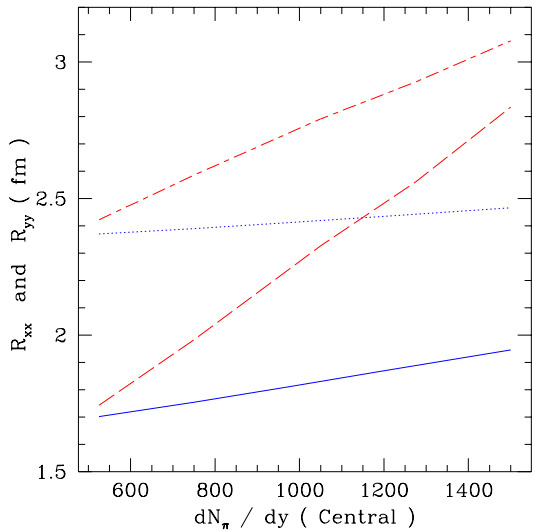}}

  \vspace{.08in}
  \caption[]
  {
   \label{hbt-fig}
     HBT radii, $R_{xx}$ and $R_{yy}$ , for $|\vec{K}|=.5 GeV$, 
     as a function of the total 
     pion multiplicity. The quoted multiplicity 
     includes all isospin states and
     is scaled by the number of participants   
     from b=8 fm to b=0 fm (a factor of 2.9) . 
     See the text for more details.
     For a resonance gas EoS $p=.2\epsilon$,  
     the solid lines and dotted lines show $R_{xx}$ and $R_{yy}$ respectively.
     For a bag model EoS, the dashed and dashed-dotted lines show 
     $R_{xx}$ and $R_{yy}$ respectively.
  }  

\end{figure}
%
\noindent 
fm).
$R_{xx}$ and 
$R_{yy}$ are shown for a bag model EoS and 
for 
a simple resonance gas EoS, $p = .2 \epsilon$.
For low energies near the left hand side of the plot, the two EoS show
approximately the same radii, roughly corresponding to the initial
elliptic shape of the matter distribution. 
For a simple resonance gas EoS, the
 HBT Radii  
show little energy dependence, while
for  an EoS with the phase transition
the homogeneity regions increase steadily with 
beam energy.
The rapid increase of 
$R_{xx}$ can be
understood qualitatively. 
The contributing pions  move in the y direction with rather high
momenta, .5 GeV. The pair therefore originates,
not from ridges of the nutshell, but from the region
in between the shells. The rapid increase of $R_{xx}$
reflects the increased separation of the nutshells at higher 
collision energies.
The increase in $R_{yy}$ reflects the flattening 
of the shells themselves. 
For flat, square-shaped, shells the
homogeneity regions are larger than for curved elliptic shells.

%

8. In conclusion, for non-central heavy ion collisions, we predict
an unusual space time evolution, which 
results from the interplay
of a hard and soft EoS typical 
of the QCD phase transition. We
have called this the ``nutcracker" flow since two shells are
produced and then separate. The azimuthal momentum asymmetry can
be seen in the higher harmonics and flow of heavy secondaries, while
the spatial asymmetry can be seen in HBT interferometry.

We end with the experimental strategy. As the ``nutcracker" flow
persists for all sufficiently non-central events, and because the
principal RHIC detectors can determine the impact parameter plane,
absolutely $any$ observable, from strangeness, to flow, to $J/\psi$ suppression,
should display marked azimuthal dependence, which reflects
the fireball in different stages. We urge our experimental colleagues
to look for this dependence from the first day of operation
at RHIC.

{\bf Acknowledgments}.
We thank J. Pons for essential numerical advice 
during the initial stages of this
work and H. Sorge  for many interesting discussions. 
This work is partially
supported by US DOE, by the grant No. DE-FG02-88ER40388.
%
%


\end{narrowtext}
\end{document}